\documentclass[12pt,preprint,epsf]{aastex}

\shorttitle{Eccentricity of the Moon's Orbit}
\shortauthors{Krisciunas} 

\begin{document}
\received{21 December 2009}

\title{Determining the Eccentricity of the Moon's Orbit without a Telescope,
and Some Comments on ``Proof'' in Empirical Science}
\author{
Kevin Krisciunas\altaffilmark{*}
}
\altaffiltext{*}{George P. and Cynthia Woods Mitchell Institute for Fundamental 
Physics \& Astronomy, Texas A. \& M. University, Department of Physics,
  4242 TAMU, College Station, TX 77843; {krisciunas@physics.tamu.edu} }

\begin{abstract} 

Prior to the invention of the telescope many astronomers worked out theories
of the motion of the Moon.  The purpose of such theories was to be able
to predict the {\em position} of the Moon in the sky.  These geometrical models
implied a certain range of distance of the Moon.  Ptolemy's most quoted model, in fact,
predicted that the Moon was nearly twice as far away at apogee than at perigee.
Measurements of the angular size of the Moon were within the capabilities of
pre-telescopic astronomers.  These could have helped refine the models of
the motion of the Moon, but hardly anyone seems to have made any measurements
that have come down to us.
Using a piece of cardboard with a small hole punched in it which
slides up and down a yardstick, we show that it is possible to determine an approximate value
of the eccentricity of the Moon's orbit.  On the basis of 64 observations taken
over 14 cycles of the Moon's phases we find $\epsilon \approx$ 0.041
$\pm$ 0.004.  A typical measurement uncertainty of the Moon's angular size
is $\pm$ 0.7 arcmin.  Since the Moon's angular size ranges from 29.4 to 33.5
arcmin, carefully taken naked eye data are accurate enough to demonstrate
periodic variations of the Moon's angular size.

\end{abstract}


\section{Introduction}

The {\em small angle formula} is one of the simple geometrical relationships we
discuss in elementary astronomy classes.  An object of diameter $d$ viewed
at distance $D$ will subtend an angle $\theta = d/D$ in radians if $D >> d$.
Two objects that satisfy this criterion are the Sun and Moon.

The history and significance of estimating the angular diameter of the Moon is a curious 
one. We briefly review it here.  Aristarchus of Samos (ca. 310$-$230 BC) wrote a treatise 
called {\em On the Sizes and Distances of the Sun and Moon}, which has 
survived.\footnote[1]{Thomas Heath, {\em Aristarchus of Samos: the Ancient Copernicus} 
(Clarendon Press, Oxford, 1913), pp. 351-414.} In it Aristarchus states that the Sun and 
Moon must have about the same angular size, because during a total solar eclipse the Moon 
just barely covers the Sun.  He also states that the angular size of the Moon is 2 degrees.  
In the {\em Sand Reckoner} Archimedes (ca. 287$-$212 BC) says that Aristarchus also used 
another value for the angular size of the Moon,\footnote[2]{Heath, note 1, pp. 353, 383.} 
1/720th part of a circle, or 0.5 deg.\footnote[3]{Albert Van 
Helden, {\em Measuring the Universe: Cosmic Dimensions from Aristarchus to Halley} 
(Chicago and London: University of Chicago Press, 1985), pp. 8-11.} This is close to the 
true mean value of 0.5182 deg = 31$^{\prime}$ 5.\hspace{-1.0 mm}$^{\prime 
\prime}$5.\footnote[4]{Arthur N. Cox, ed. {\em Allen's Astrophysical Quantitities} 
(Springer-Verlag, New York, 2000), pp. 308-309.  Two times the ``mean Moon radius'' (1738.2 
km) divided by the mean distance (384,401 km), times 180/$\pi$ = 0.5182 deg = 31.09 arcmin.  
Given the extreme range of the Moon's distance (356,400 to 406,700 km), its geocentric 
angular size ranges from 29.39 to 33.53 arcmin. The mean distance of the Moon divided by 
the equatorial radius of the Earth, 6378 km, gives a mean geocentric distance of 60.27 
Earth radii.  Note that the maximum distance is 5.8 percent greater than the mean distance, 
and the minimum distance is 7.3 percent less than the mean distance.}

Hipparchus (ca. 190$-$120 BC) also wrote a treatise on the sizes and distances of the Sun 
and Moon.  This has not survived, but its contents are known from two 
sources.\footnote[5]{G. J. Toomer, ``Hipparchus on the distances of the Sun and Moon,'' 
Archive for the History of Exact Sciences, {\bf 14}, 126--142 (1975).}  Hipparchus obtained 
a mean lunar distance of 67$\frac{1}{3}$ Earth radii, with a range of 62 to 
72$\frac{2}{3}$. The range is $\pm$7.9 percent. In the {\em Almagest} (IV,9) Ptolemy (ca. 
100--170 AD) quotes Hipparchus's value for the Moon's mean angular size of 1/650th of a 
circle, or 33$^{\prime}$ 14$^{\prime \prime}$.\footnote[6]{G. J. Toomer, {\em Ptolemy's 
Almagest} (Berlin, Heidelberg, Tokyo: Springer-Verlag, 1984), p. 205.}

Ptolemy's values for the minimum and maximum angular size of the Moon were 31$^{\prime}$ 
20$^{\prime \prime}$ and 35$^{\prime}$ 20$^{\prime \prime}$, 
respectively.\footnote[7]{Toomer, note 6, pp. 254, 284.}  But Ptolemy's model for the {\em 
position} of the Moon in the sky implies that its minimum distance is a mere 33.55 
Earth radii, while its maximum distance is 64.17 Earth radii.\footnote[8]{Toomer, note 
6, pp. 251, 259.  In section 5.13 of the {\em Almagest}
Ptolemy gives the mean distance of the Moon at syzygy (i.e. new/full Moon) of     
59 Earth radii.  The mean distance at quadrature (i.e. first or third quarter) 
is 38$\frac{43}{60}$ Earth radii, and
the radius of the epicycle is 5$\frac{10}{60}$ Earth radii.  It follows that
the greatest distance occurs at syzygy and is 59 + 5$\frac{10}{60}$ = 64.17 Earth
radii.  The minimum distance occurs at quadrature and is
38$\frac{43}{60}$ minus 5$\frac{10}{60}$  = 33.55 Earth radii.  See also note
17 below.}  Given the importance of the {\em Almagest} from ancient times until the Renaissance, the 
implication that the Moon's distance (and, hence, angular size) ranges by nearly a factor 
of two was known to all those who worked on models of the Moon's motion, though any serious 
observer of the Moon would have known that its range of angular size is considerably 
smaller. 

Simplicius (6th century AD), in his commentary on Aristotle's {\em De Caelo}, wrote:
``...if we observe the moon by means of an instrument ... it is at one time a disk of
eleven finger-breadths, and again at another time a disk of twelve finger-breadths.''
Here a ``finger-breadth'' (digit, or {\em daktylos}) is {\em not} the angular size
of a finger at some distance.  It is a Babylonian unit equal to
one-twelfth of a {\em degree}, but in this context it might be one-twelfth of the
maximum angular size of the Moon.  Taken at face value, the statement by Simplicius
implies the existence of observations that give just about the right range of the Moon's
angular size.  But the data and the identity of the observer are not 
given.\footnote[9]{Thomas L. Heath, {\em Greek Astronomy} (Dover, New York, 1991), pp. xvii, 69.
See also the anonymous article, ``Babylonian measures and the {\em daktylos},''
Observatory, {\bf 42}, 46--51 (1919).}

Finally, in the fourteenth century two astronomers showed an interest in actually making 
some measurements. Levi ben Gerson (1288-1344), also known as Gersonides, was a considerably 
versatile and 
accomplished scholar.\footnote[10]{Bernard R. Goldstein, {\em The Astronomy of Levi ben 
Gerson (1288--1344)} (New York: Berlin, Heidelberg, Tokyo: Springer-Verlag, 1985).} He 
invented the staff of Jacob, which consists of a calibrated ruler that slides 
perpendicularly along another calibrated staff.  With it one could determine the angular 
separation of two stars in the sky or determine the height of building.  Using a calibrated 
staff and a pinhole camera, he determined that the Sun ranges in angular size from 
27$^{\prime}$ 50$^{\prime \prime}$ 
to 30$^{\prime}$ 0$^{\prime \prime}$.\footnote[11]{Goldstein, note 10, p. 113.} He 
commented on Ptolemy's factor-of-two range of the Moon's implied angular size. Using the 
staff and pinhole camera, Levi found a lunar diameter at quadrature only slightly larger 
than at opposition (i.e. full Moon).\footnote[12]{Goldstein, note 10, pp. 105, 186.} He did 
{\em not} carry out observations at all lunar phases.

Another pre-telescopic astronomer to comment specifically on the range of lunar angular 
size as a consequence of a model to explain its varying position was Ibn al-Shatir 
(1304--1375/6 AD).\footnote[13]{V. Roberts, ``The solar and lunar theory of Ibn ash-Shatir: 
a pre-Copernican model,'' Isis, {\bf 48}, 428--432 (1957).}  The implied range was 
29$^{\prime}$ 2$^{\prime \prime}$ to 37$^{\prime}$ 58$^{\prime \prime}$.

Regiomontanus (1436$-$1476) doubted the very large range of lunar angular size implied by 
Ptolemy's model, basing his criticism on al-Battani's apparent diameters at 
syzygy.\footnote[14]{Johannes Regiomontanus, {\em Epytoma in almagesti Ptolemei} 
(Venice, 1496), \S 5.22. There is no published translation of this work.
al-Battani lived from about 858 to 929 AD.}  In {\em De Revolutionibus} Copernicus 
(1473$-$1543) gives a range of 30$^{\prime}$ to 35$^{\prime}$ 38$^{\prime \prime}$ for the 
Moon's angular size.\footnote[15]{Edward Rosen, {\em Nicholas Copernicus: On the 
Revolutions} (Baltimore and London: Johns Hopkins Univ. Press, 1992), p. 209.}

Tycho Brahe (1546$-$1601), the greatest of all pre-telescopic astronomers, devoted 
considerable effort to the orbit of the Moon, but he was concerned only with the Moon's 
varying position.  Still, his model implied a range of lunar distance of 5.8 percent at 
syzygy,\footnote[16]{N. Swerdlow, ``The lunar theories of Tycho Brahe and Christian 
Longomontanus in the {\em Progymnasmata} and {\em Astronomia Danica}, Annals of Science, 
{\bf 66}, 5--58 (2009), on p. 43.} or $\pm$~2.9 percent.

Of course, we now can model the orbit of the Moon quite accurately.  But we should not 
simply say that the eccentricity of the Moon is $\epsilon$ = 0.05490,$^4$ implying a range 
of distance of $\pm$ 5.5 percent.  Contrary to what introductory astronomy textbooks say, 
the Earth does not orbit the Sun on a simple elliptical orbit, and the Moon does not orbit 
the Earth on a simple elliptical orbit either.  The Earth-Moon barycenter orbits the Sun on 
a nearly Keplerian ellipse.  The Earth and Moon do something else.  Let $\delta$ = 0.011 be 
the ``amplitude of Ptolemy's evection.'' To first order the maximum deviation from uniform 
angular motion of the Moon varies from $\pm$2($\epsilon - \delta $) when the Moon is new or 
full and $\pm$2($\epsilon + \delta $) for first and third quarters.  The distance of the 
Moon (to first order) varies by ($\epsilon + \delta$) = $\pm$6.6 percent.\footnote[17]{M. 
C. Gutzwiller, ``Moon-Earth-Sun: the oldest three-body problem,'' Reviews of Modern 
Physics, {\bf 70}, 589-639 (1998), pp. 601-602.  Compared to uniform motion
against the background of stars, the new/full Moon can be 5 deg ahead or behind.  At
first/third quarter the Moon can be 7.5 deg ahead or behind the mean motion.  To 
quote Gutzwiller: ``This new feature is known as the {\em evection}.
Ptolemy found a mechanical analog for this
peculiar complication, called the crank model.  It describes the angular coupling
between the Sun and Moon correctly, but it has the absurd consequence of causing the
distance of the Moon from the Earth to vary by almost a factor of 2.''} Three modern 
ephemerides use 669, 921, and 915 terms, respectively, to calculate the 
distance to the Moon!\footnote[18]{Gutzwiller,  note 17, p. 628.}

One might think that Hipparchus or some other pre-telescopic astronomer
had actually measured the variation of the
apparent angular size of the Moon over a number of lunations.  Even if this is true,
no data set or analysis of one has come down to us.  With an instrument
as simple as a quarter-inch diameter sighting hole viewed at some distance down a
yardstick, is it possible to demonstrate that the Moon's angular size varies
from 29 to 33 arcmin in a quasi-sinusoidal way?

\section{Observations}

Hipparchus measured the angular size of the Moon using a {\em dioptra}.  Such a device uses 
a round object of small angular size to {\em occult} the Moon.  As did Levi ben 
Gerson,\footnote[18]{Goldstein, note 10, p. 156.} we have found that it is better to use a 
sighting hole.

We fashioned a cross piece of cardboard that can slide up and down a yardstick.  The
yardstick is calibrated in centimeters on one edge. There are tick marks every 5 mm.
The purpose of the cross piece is simply to hold a thin piece of cardboard that has a
small hole punched through it.  We estimate that the diameter of the sighting hole is
6.2 mm.

In Table~\ref{data} we present a series of observations carried out during 2009. All 
the data were taken at my home in College Station, Texas, which is at an elevation of about 
100 m above sea level.  All observations were made with my left eye, which is my better 
eye.  In Table~\ref{data} the age of the Moon is the number of days since the previous 
new Moon.\footnote[19]{aa.usno.navy.mil/data/docs/MoonPhase.php\#y2009} The ``true angular 
size'' is the geocentric angular diameter interpolated from values given in the {\em 
Astronomical Almanac}.\footnote[20]{{\em The Astronomical Almanac for the Year 2009} 
(Washington, D. C., Nautical Almanac Office, 2009).} Observations from 11:15 to 13:15 UT 
and from 22:37 to 23:43 UT were made during twilight or daytime.  This constituted a 
majority of the observations.  Starting on 04 August 2009 our data values in column 5 of 
Table~\ref{data} are typically the average of two measurements, the first obtained 
while moving the cross piece {\em out} from a position much too close to the eye, the 
second obtained by moving the cross piece {\em in} from the far end of the yardstick.  A 
pair of such observations made on the same occasion typically exhibits a difference of 8 mm.  
So a typical internal error of one of the averages in column 5 is $\pm$ 4 mm.

Let $D$ be the distance down the yardstick that the 6.2 mm sighting hole is viewed.
Using the small angle approximation one obtains the observed angular size of the
Moon as follows:
\begin{equation}
\theta _{\rm {obs}}(\rm {arcmin}) \; = \;  \left( \frac{6.2}{D}\right)  \; \times \;
 \left(\frac{180}{\pi}\right)  \; \times \ 60 \; \; .
\end{equation}

Taken at face value, our mean angular size is 25.9 arcmin.  Thus, our preliminary
data exhibit a systematic error of 5.2 arcmin.  However, consider that the pupil has
a non-zero size.  In fact, the pupil's size is comparable to the size of the sighting
hole.

We therefore devised a simple calibration.  A disk of diameter 90.44 mm would subtend an
angle of 31.09 arcmin if situated at a distance of 10 m.  We cut out a disk measured to
be 91 mm in diameter.  Situated at 10 m, the angular size was 31.28 arcmin.  If it were
not for the finite size of the pupil, a 6.2 mm hole viewed at 681.3 mm would subtend an
equal angle.  If we place the sighting hole at this distance,
we can see beyond the left side of the disk from the right side of the pupil, and we
can we beyond the right side of the disk from the left side of the pupil.  So we
need to place the sighting hole at a {\em greater} distance from the eye to make
the best apparent fit of the disk as seen through the hole.
I found that I had to place the sighting hole 821 mm from my left
eye to produce the best match to the angular size of the 91 mm disk.  (This calibration
was done at the office on 5 August 2009 in an illuminated hallway.)  We thus obtain a
correction factor of 821/681.3 = 1.205, which should be used to scale our preliminary
angular sizes.  In other words, from April through November 2009 when these
lunar observations were made, for my eye
\begin{equation}
\theta _{\rm {corr}} \; = \; \theta _{\rm {obs}} \; \times \; 1.205 \; .
\end{equation}
Further justification of the correction factor is shown in  Fig.~\ref{ratio},
in which we plot the ratios of true (geocentric) angular sizes to the corresponding
observed values.  One outlier is eliminated from analysis.
A simple regression line gives a non-zero slope at the 2.3-$\sigma$
level of significance.  Thus, there is marginal (but not statistically significant)
evidence that I was measuring the Moon larger over time.  The average value of the
ratios is 1.199 $\pm$ 0.005, which is very close to the correction factor derived
from a measurement of the 91 mm disk at 10 m.  Ideally, one would do this calibration
after each measurement of the Moon under lighting conditions as similar as
possible.  Maybe there is a sufficiently good set of three correction factors for each
observer, one for nighttime, one for twilight, and one for daytime observations.
But this is beyond the scope of the present paper.

We note that the correction factor eliminates any systematic error in the
adopted size of the sighting hole.  If the true diameter of the sighting
hole were really 6.4 mm, then the correction factor would be correspondingly
smaller.  For the calibration given above and observations with my left eye, with
minimal error for the Moon's varying angular size I could use the following:
\begin{equation}
\theta _{\rm {corr}} \; = \; \left(\frac{821}{D}\right) \; \times \; 31.28 \;  .
\end{equation}
Thanks to the correction factor, our mean value for the angular size is
31.18 arcmin, very close to the true mean value of 31.09.

\section{Analysis and Discussion}

Even before the ancient Greeks, the Babylonians knew the anomalistic period
of the Moon.  This is the time between occurrences of maximum daily motion against
the background of stars (i.e. the time from perigee to perigee),
27.55455 days.$^4$ We have taken data over 7 lunations = 7.5 anomalistic
months.  Ideally, one would have a data set that extends over 14 lunations
= 15 anomalistic months.  In that case every phase would occur at some time
over the range of observations at perigee and at apogee.

Since all observations presented in Table~\ref{data} were made by the same observer
with the same eye and the same sighting hole, we could derive a value of the
eccentricity of the Moon's orbit from an analysis of the distances-down-the-yardstick
given in column 5 of the table, or we could use the uncorrected (observed) angular
sizes of the Moon.  The eccentricity of the orbit is just the amplitude of the sinusoid
that best fits the data, divided by the mean value of the data.  But we choose to use
the corrected angular sizes given in column 6 of Table~\ref{data} because the
use of the correction factor eliminates a significant source of systematic error.

We note that while a Keplerian elliptical orbit is necessarily eccentric, without more
accurate data than we present here we cannot prove that the Moon's orbit is
elliptical, circular, ovoid, or some other shape.  We have only demonstrated that
the Earth does not reside at the center of the Moon's orbit, as was known
by Ptolemy, Copernicus, and others.

Say we knew nothing of the value of the anomalistic month.  What does our data set tell us 
is the time between perigees?  We have eliminated one data point from the analysis, the 
first datum from October 23rd, which is a 4.5-$\sigma$ outlier.  Using a program for the 
analysis of variable star light curves,\footnote[21]{M. Breger, ``PERDET: multiple PERiod 
DETermination User Manual,'' Communications in Astroseismology, No. 6 (1989).} we obtain a 
best fit period of 27.24 $\pm$ 0.29 days.  The uncertainty of the period is to be 
considered a lower limit.\footnote[22]{M. H. Montgomery, D. O'Donoghue, ``A derivation of 
the errors for least squares fitting to time series data,'' Delta Scuti Star Newsletter, 
No. 13, 28-32 (1999).} The amplitude of the variation is 1.23 $\pm$ 0.17 arcmin.  
The root-mean-square scatter of the data with respect to that sinusoid is $\pm$ 0.74
arcmin, which is the internal random error of the measurements. The 
implied eccentricity of the Moon's orbit is 0.039 $\pm$ 0.006. In the top graph of 
Fig.~\ref{ang_diam} we show our data folded with the best period derivable from only our 
data set.

The bottom graph of Fig.~\ref{ang_diam} shows the geocentric angular size of the Moon (from 
the {\em Astronomical Almanac}) at the dates and times we took data, folded with the 
anomalistic period of the Moon. Note that the actual ``true values'' do not exhibit a 
simple sinusoid.

Comparison of our corrected angular sizes to the geocentric values from
the {\em Astronomical Almanac} indicates that an individual data point
of ours is accurate to $\pm$ 0.80 arcmin.  This is only marginally larger than
the internal random error, implying that the corrected angular sizes contain no
serious systematic offset.  One way to slightly improve the analysis would
be to correct the ``true'' angular sizes based on the geocentric distance of
the Moon to the topocentric values.

The data obtained when the
age of the Moon is between 7.7 and 22.4 days (i.e. first quarter to
third quarter) are a bit more accurate than observations of the crescent Moon.
We believe that twilight and daytime observations are more accurate than nighttime
observations, because the Moon's glare makes the observations more difficult.

It is remarkable that none of the great pre-telescopic astronomers carried
out simple observations like those presented here.  Of course, before the
19th century there was no Fourier analysis or least-squares theory to derive
some of the numbers presented here, but the Babylonians {\em did} know
the length of the anomalistic month to a high degree of accuracy, so measurements
of the angular diameter of the Moon could have been folded with that period using
nothing more than simple arithmetic.  It could be asserted that because no known
pre-telescopic data set has come down to us, the analysis presented here is
completely unhistorical, and therefore of no interest to historians.  Perhaps
this paper will motivate scholars who can read Arabic or Hebrew to identify and
scrutinize some previously unstudied manuscripts that contain data like those presented
here.

Granted, the geometrical models of the Moon's motion were concerned with its
ecliptic latitude and longitude, not with its distance.  But observations of the
Moon's angular size, which is to say measures related to its physical distance,
could have nudged astronomers down the path of more physically realistic models
of the solar system.

\vspace {1 cm}

\acknowledgments

I thank Dudley Herschbach for suggesting that I write this article.
I am  particularly grateful to Noel Swerdlow for numerous fruitful discussions.
Alexei Belyanin, Gerald Handler, and an anonymous referee provided useful
references.  I also thank James Evans and Stephen O'Meara for useful comments.

\appendix

\section{Additional Data and Some Commentary}

In this appendix we provide some additional data obtained since 
mid-November, 2009. We also give some philosophical comments not 
included in the paper published in the {\em American Journal of 
Physics} in August of 2010.

In Table~\ref{more_data} we give an additional 29 measurements of the 
Moon. The average ratio of the true (geocentric) angular size to the 
observed angular size is 1.158 $\pm$ 0.006 for these measures.  
Eliminating the one outlier from 23 October 2009, the average ratio
of ``true'' to ``observed'' angular size is 1.18 for 64 good measurements.

Our data set now covers 14 cycles of the Moon's phases and 15 cycles of
its anomalistic period.  Thus, all waxing and waning phases of the
Moon occurred at apogee and perigee over this time period.   

In Table~\ref{corr_factor} we give seven measures of the correction
factor C.  The first three were made during the daytime in lighted
hallways at Texas A\&M University.  The first and third measures
were made at the same location, but the measure of 27 January 2010
was made elsewhere.  The final four measures were made during
twilight at the author's home, in an attempt to calibrate the
cross staff under conditions as similar as possible to many of the
observations of the Moon made during twilight.

It seems that my correction factor was not a constant over the
course of the 14 months of these observations.  Exactly why I
cannot say, but at age 56 my eyes are not what they once were.  Perhaps
my left eye was changing measurably over the past year, given this evidence.
It has been suggested that maybe the way to get around the potential of a
variable correction factor is to use a 1 mm sight (or smaller) at the 
eye end of the yardstick.  

What, then, is the best way to demonstrate the sinusoidal variation of
the angular size of the Moon?  In the world of variable star observing
one could have a secular variation of brightness of a star and cyclical
changes on top of that.  For our whole data set we fit a fourth
order polynomial to the uncorrected angular sizes, then subtracted that
curve.  This would take care of any slow trends in the actual correction
factor over time.  The mean value of $\theta _{obs}$ is 26.38 arcmin.
Since the true mean angular size of the Moon is 31.09 arcmin, that is
also confirmation that our true mean correction factor must be about 1.18.

In Fig. \ref{ang_diam_64} we have eliminated one outlier (the first 
observation of 23 October 2009) and have folded the residuals
of $\theta _{obs}$ with a 
revised period of 27.5854 $\pm$ 0.0955 d.  Note that this is only 0.03
d longer than the true known anomalistic period of the Moon.  We find
an epoch of apogee (i.e. minimum angular size) of Julian Date 2,454,965.8794,
or 14 May 2009 at 9 hours UT.  This is precisely one date of minimum
angular size given in the 2009 volume of the {\em Astronomical Almanac}.

Based on 64 measurements we find the eccentricity of the Moon's orbit to
be $\epsilon$ = (1.0853 $\pm$ 0.109)/26.38 = 0.041 $\pm$ 0.004, just about
what we obtained from the first 35 good data points.  On the basis of our
naked eye observations, the corrected amplitude of the sinusoid that describes the  
variation of the Moon's angular size is A = 1.18 $\times$ 1.0853 $\approx$
1.28 arcmin.  The average random error of the 64 measurements is 
$\pm$ 0.72 arcmin.  The first measurement from 23 October 2009 can be
shown to be a 3.3-$\sigma$ outlier even without any information of the
true angular size of the Moon.

Using the methods described in this paper, students in my Astronomy 101
class made 34 measurements of the Moon on 22 February 2010 UT.  The 
median value was 31.35 arcmin, exactly equal to the true angular size
on that night!  However, the standard devation of their measures was
$\pm$ 4.1 arcmin.  I suspect that a novice observer could not measure
the variation of the Moon's angular size very well.  The internal
random error has to be less than 1.0 arcmin to succeed.  Even I practiced
for 4 months making observations with my worse eye until I realized I
had to start over from scratch using my better eye.  And I found that
twilight observations are usually the best.  In the upper graph
of Fig. \ref{ang_diam_64} the triangles make a tighter sinusoid than the dots.

One of the motivations of this paper was to provide an example to our 
students of the scientific method.  As outlined in many an astronomy 
text,\footnote[23]{See, for example: Jeffrey Bennett, Megan Donahue, 
Nicholas Schneider, and Mark Voit, {\em The Essential Cosmic 
Perspective}, 5th ed. (San Francisco, Addison-Wesley), 2009, p. 70 
ff.} we observe something in the laboratory or in the universe and we 
try to explain what we see.  A good hypothesis allows us to make 
predictions.  We can then make some measurements.  If the data agree 
with the predictions, that helps confirm the hypothesis.  If the data 
do not agree with the predictions, then the data might be wrong 
somehow, maybe the hypothesis needs modification, or maybe we need a 
completely new hypothesis.

My experience is that when students hear about Kepler's Laws of 
planetary motion they only try to memorize them for upcoming quizzes 
and tests.  These laws are quite abstract to the average 
non-astronomer.  If we can get the students to confront some data, 
that should allow them an understanding at a deeper level.

So here is one hypothesis: though the Moon is not a planet, it orbits
the Earth (or, more exactly, the Earth-Moon barycenter) on an elliptical
orbit having an eccentricity of a few percent.  If we measure the
angular size of the Moon, and the data show no statistically significant
changes of the Moon's size, then the hypothesis might be wrong, or the
data might not be accurate enough to demonstrate the effect.  If the
data show regular $\pm$ 4 percent variations of the Moon's angular size,
like our data in Tables~\ref{data} and \ref{more_data} and shown in
Figs.~\ref{ang_diam} and \ref{ang_diam_64}, then under the assumption that the
Moon's {\em linear} size is constant (i.e. that the Moon does not pulsate), we
have evidence that the Moon's distance from the Earth varies $\pm$ 4 percent.
I assert that this is {\em confirming evidence} that Kepler's
First Law holds for the Moon.  As we state in the third paragraph 
of \S3 above, our data are not accurate enough to prove that the
orbit is elliptical, circular, ovoid, or some other shape.  But we
have {\em proven} that the Earth is not at the center of the Moon's
orbit.  

The {\em American Journal of Physics} uses two referees for each paper.
One of our original two referees said that the first submitted version
of this paper was acceptable for publication.  The second referee was
an historian and had some serious reservations.  The second referee
absolutely would not allow me to say that we have confirming evidence
for Kepler's First Law of orbital motion.  That referee and then a
third referee would not let me state that the data {\em prove} that
the Earth is offset from the center of the Moon's orbit.  I was required
to use either of the verbs ``show'' or ``demonstrate''.

Admittedly, the words ``prove'' and ``proof'' are seldom used in 
scientific papers.\footnote[24]{See: Douglas Clowe, Maru\v{s}a 
Brada\v{c}, Anthony H. Gonzalez, Maxim Markevitch, Scott W. Randall, 
Christine Jones, and Dennis Zaritsky, ``A direct empirical proof of 
the existence of Dark Matter,'' {\em Astrophysical Journal}, {\bf 
648}, L109--L113 (2006).  One time I really wanted to use the word 
``proof'' but opted for ``strong evidence''. See: Kevin Krisciunas, R. 
F. Griffin, E. F. Guinan, K. D. Luedeke, and G. P. McCook, ``9 
Aurigae: strong evidence for non-radial pulsations,'' {\em Monthly 
Notices of the Royal Astronomical Society}, {\bf 273}, 662--674 
(1995).  We {\em did} prove the existence of a new class of 
non-radially pulsating stars.  See also: Anthony B. Kaye, Gerald 
Handler, Kevin Krisciunas, Ennio Poretti, and Filippo M. Zerbi, 
``$\gamma$ Doradus stars: defining a new class of pulsating 
variables,'' {\em Publications of the Astronomical Society of the 
Pacific}, {\bf 111}, 840--844 (1999).}  And of course, in an
empirical science such as astronomy, we would not use it in the
same way as in a mathematical or geometrical proof, which must 
be absolutely true given certain starting assumptions.
Still, in astronomy, we can say we have proven something.
Much more common, and less 
strong, is to say that we have ``evidence for'' something, or that the 
data are ``consistent with'' some hypothesis.  A still weaker claim is 
to say that the data are ``not inconsistent with'' some hypothesis. 
For example, if two measures should be the same but differ by three 
standard deviations, then we might say that the two results are not 
inconsistent with each other. We feel more confident with the 
agreement if the two numbers are within one standard deviation.  Then 
we say that the two numbers are statistically equal.

Under the {\em assumption} that a single sine wave fits the
variations of our corrected angular sizes of the Moon,
we did not quite find $\epsilon$ = 
0.055.  Owing to the joint gravitational forces of the Earth and Sun, 
the actual orbit of the Moon is quite complicated.  Its minimum 
distance is 7.3 percent closer than the mean distance, and its maximum 
distance is 5.8 percent greater than the mean distance.  But we did 
prove/show/demonstrate that pre-telescopic astronomers could have 
generated a data set like ours presented here.  It would be 
interesting to investigate when ``consistency checks'' became a part 
of the modern method of doing science.  Had the pre-telescopic 
astronomers used angular size measurements to make consistency checks 
of their models of the Moon's motion, they would have easily shown 
that Ptolemy's most quoted model could not be physically correct. One 
main point of the Copernican revolution, as I understand it, is that 
by the end of the 16th century we were ready to go beyond geometrical 
models that merely accounted for the varying positions of the Moon, 
Sun, and planets against the background of stars.  We were ready to 
determine the true, physical arrangement of objects in the solar 
system.

\newpage

\begin{deluxetable}{cccrccc}
\tablewidth{0pc}
\tablecaption{Lunar Data\label{data}}
\tablehead{   \colhead{2009 UT Date} & \colhead{UT} & \colhead{JD\tablenotemark{a}} &
\colhead{Age\tablenotemark{b}} & \colhead{D\tablenotemark{c}} & 
\colhead{$\theta _{corr}$} & \colhead{$\theta _{true}$\tablenotemark{d}}  } 
\startdata
21 Apr &   11:23 &  4942.9743  &     25.80 &  874.0 &  29.39 & 30.69 \\
06 May &   03:55 &  4957.6632  &     11.02 &  840.0 &  30.58 & 31.03 \\
29 May &   01:33 &  4980.5646  &      4.56 &  810.0 &  31.71 & 32.52 \\
31 May &   01:33 &  4982.5646  &      6.56 &  794.0 &  32.35 & 31.82 \\
14 Jul &   11:15 &  5026.9688  &     21.65 &  828.0 &  31.02 & 30.65 \\
16 Jul &   12:02 &  5029.0014  &     23.69 &  812.0 &  31.63 & 31.65 \\
29 Jul &   01:33 &  5041.5646  &      6.96 &  816.0 &  31.47 & 30.62 \\
04 Aug &   02:08 &  5047.5889  &     12.98 &  859.0 &  29.90 & 29.42 \\
07 Aug &   12:01 &  5051.0007  &     16.39 &  846.5 &  30.34 & 29.72 \\
10 Aug &   11:58 &  5053.9986  &     19.39 &  846.5 &  30.34 & 30.40 \\
13 Aug &   12:04 &  5057.0028  &     22.40 &  796.0 &  32.27 & 31.48 \\
14 Aug &   11:57 &  5057.9979  &     23.39 &  817.0 &  31.44 & 31.90 \\
15 Aug &   12:08 &  5059.0056  &     24.40 &  816.0 &  31.47 & 32.31 \\
17 Aug &   10:00 &  5060.9167  &     26.31 &  805.0 &  31.90 & 32.96 \\
27 Aug &   02:48 &  5070.6167  &      6.70 &  810.0 &  31.71 & 30.26 \\
30 Aug &   03:00 &  5073.6250  &      9.71 &  870.5 &  29.50 & 29.70 \\
02 Sep &   01:12 &  5076.5500  &     12.63 &  899.0 &  28.57 & 29.56 \\
07 Sep &   02:48 &  5081.6167  &     17.70 &  811.0 &  31.67 & 30.83 \\
09 Sep &   11:45 &  5083.9896  &     20.07 &  795.0 &  32.31 & 31.23 \\
16 Sep &   11:46 &  5090.9903  &     27.07 &  804.0 &  31.94 & 32.80 \\
27 Sep &   23:30 &  5102.4792  &      9.20 &  848.0 &  30.29 & 29.54 \\
29 Sep &   23:42 &  5104.4875  &     11.21 &  865.0 &  29.69 & 29.68 \\
03 Oct &   03:51 &  5107.6604  &     14.38 &  859.5 &  29.88 & 30.41 \\
08 Oct &   03:40 &  5112.6528  &     19.37 &  841.0 &  30.54 & 31.73 \\
08 Oct &   13:00 &  5113.0417  &     19.76 &  802.0 &  32.02 & 31.81 \\
15 Oct &   11:54 &  5119.9958  &     26.72 &  761.5 &  33.73 & 32.23 \\
23 Oct &   01:27 &  5127.5604  &      4.83 &  765.0 &  33.57 & 29.95 \\
23 Oct &   23:29 &  5128.4785  &      5.75 &  840.0 &  30.58 & 29.74 \\
24 Oct &   23:42 &  5129.4875  &      6.76 &  843.0 &  30.47 & 29.60 \\
25 Oct &   22:37 &  5130.4424  &      7.71 &  868.0 &  29.59 & 29.56 \\
27 Oct &   23:43 &  5132.4882  &      9.76 &  819.0 &  31.36 & 29.75 \\
31 Oct &   23:28 &  5136.4778  &     13.75 &  817.0 &  31.44 & 30.98 \\
03 Nov &   12:51 &  5139.0354  &     16.30 &  792.0 &  32.43 & 31.85 \\
05 Nov &   12:42 &  5141.0292  &     18.30 &  773.0 &  33.23 & 32.26 \\
06 Nov &   12:55 &  5142.0382  &     19.31 &  789.5 &  32.53 & 32.35 \\
10 Nov &   13:15 &  5146.0521  &     23.32 &  800.0 &  32.10 & 32.11 \\
\enddata
\tablenotetext{a}{Julian Date {\em minus} 2,450,000.}
\tablenotetext{b}{Time in days since previous new Moon.}
\tablenotetext{c}{Distance from the eye (in mm) of the 6.2 mm diameter sighting hole.
Typical internal error is $\pm$ 4 mm.}
\tablenotetext{d}{Geocentric angular diameter of the Moon (in arcmin), 
interpolated from data given in the {\em Astronomical Almanac}.}
\end{deluxetable}

\begin{deluxetable}{cccrccc}
\tablewidth{0pc}
\tablecaption{Additional Data\label{more_data}\tablenotemark{a}}
\tablehead{   \colhead{2009/2010 UT Date} & \colhead{UT} & \colhead{JD} &
\colhead{Age} & \colhead{D} & 
\colhead{$\theta _{obs}$} & \colhead{$\theta _{true}$}  } 
\startdata
26 Nov & 23:40 & 5162.4861 & 10.18 & 791.0 & 26.95 & 30.31  \\
04 Dec & 13:05 & 5170.0451 & 17.74 & 777.0 & 27.43 & 32.85  \\
04 Dec & 14:16 & 5171.0944 & 18.79 & 760.0 & 28.04 & 32.79  \\
09 Dec & 14:11 & 5175.0910 & 22.79 & 804.0 & 26.51 & 31.79  \\
20 Dec & 23:08 & 5186.4639 & 4.46  & 840.0 & 25.37 & 29.45  \\
25 Dec & 23:37 & 5191.4840 & 9.48  & 777.0 & 27.43 & 30.65  \\
27 Dec & 23:04 & 5193.4611 & 11.46 & 786.0 & 27.12 & 31.63  \\
28 Dec & 23:20 & 5194.4722 & 12.47 & 793.0 & 26.88 & 32.16  \\
30 Dec & 23:34 & 5196.4819 & 14.48 & 748.5 & 28.48 & 33.00  \\
21 Jan & 23:38 & 5218.4847 & 6.69  & 817.5 & 26.07 & 30.16  \\
26 Jan & 23:18 & 5223.4708 & 11.67 & 781.5 & 27.27 & 32.53  \\
06 Feb & 13:57 & 5234.0812 & 22.28 & 756.0 & 28.19 & 30.67  \\
22 Feb & 00:03 & 5249.5021 & 7.88  & 763.5 & 27.92 & 31.35  \\
28 Feb & 00:24 & 5255.5167 & 13.90 & 745.0 & 28.61 & 33.38  \\
28 Feb & 04:28 & 5255.6861 & 14.07 & 784.0 & 27.19 & 33.36  \\
25 Apr & 00:49 & 5311.5340 & 10.51 & 774.5 & 27.52 & 32.53  \\
02 May & 12:16 & 5319.0111 & 17.99 & 824.0 & 25.87 & 30.42  \\
23 May & 01:07 & 5339.5465 & 9.00  & 760.0 & 28.04 & 32.15  \\
28 May & 02:44 & 5344.6139 & 14.07 & 764.0 & 27.90 & 30.99  \\
30 May & 04:38 & 5346.6931 & 16.15 & 788.0 & 27.05 & 30.34  \\
30 May & 11:10 & 5346.9653 & 16.42 & 798.0 & 26.71 & 30.26  \\
30 May & 12:08 & 5347.0056 & 16.46 & 819.0 & 26.02 & 30.25  \\
31 May & 11:42 & 5347.9875 & 17.44 & 808.0 & 26.38 & 29.98  \\
04 Jun & 11:56 & 5351.9972 & 21.45 & 838.0 & 25.43 & 29.59  \\
22 Jun & 01:16 & 5369.5528 & 9.58  & 789.0 & 27.01 & 31.42  \\
23 Jun & 01:39 & 5370.5688 & 10.60 & 790.0 & 26.98 & 31.14  \\
24 Jun & 01:41 & 5371.5701 & 11.60 & 795.0 & 26.81 & 30.87  \\
28 Jun & 11:32 & 5375.9806 & 16.01 & 835.0 & 25.53 & 29.80  \\
04 Jul & 13:12 & 5382.0500 & 22.51 & 830.0 & 25.68 & 29.92  \\
\enddata
\tablenotetext{a}{The meanings of the column labels
 are the same as in Table~\ref{data} except column
6 contains the {\em observed} (i.e., uncorrected) angular sizes, 
not the corrected ones.}
\end{deluxetable}

\begin{deluxetable}{lccc}
\tablewidth{0pc}
\tablecaption{Measures of the Correction Factor\label{corr_factor}}
\tablehead{   \colhead{UT Date} & \colhead {UT} &
\colhead{JD\tablenotemark{a}} & \colhead{C}   } 
\startdata
05 Aug 2009 & \ldots & 5049.25 & 1.205 \\
27 Jan 2010 & \ldots & 5224.33 & 1.151 \\
29 Jan 2010 & \ldots & 5226.31 & 1.118 \\
06 Feb 2010 & 14:07  & 5234.09 & 1.118 \\
05 May 2010 & 01:05  & 5321.55 & 1.083 \\
28 Jun 2010 & 11:39  & 5375.98 & 1.072 \\
06 Jul 2010 & 01:34  & 5383.56 & 1.086 \\

\enddata
\tablenotetext{a}{Julian Date {\em minus} 2,450,000.}
\end{deluxetable}

\clearpage

\figcaption[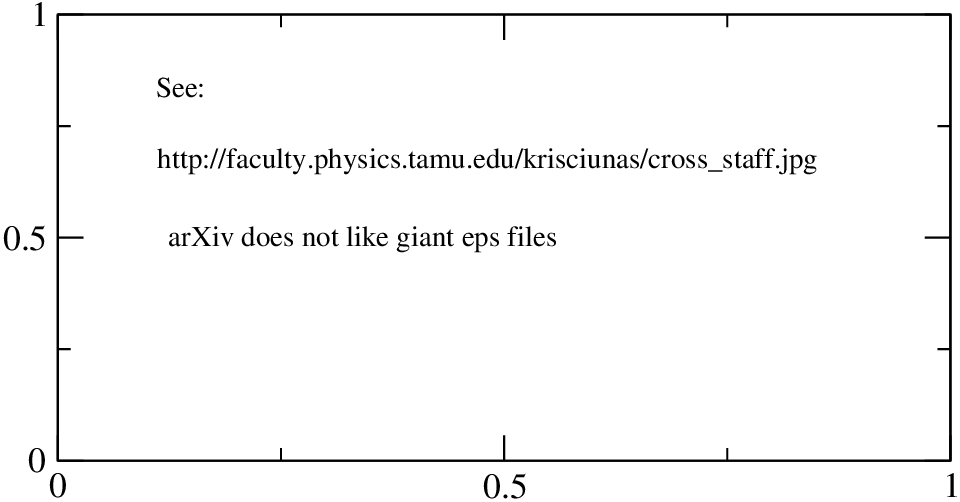]{A simple Moon measuring device, made
from the bottom of a box of pancake mix. \label{cross_staff}
}

\figcaption[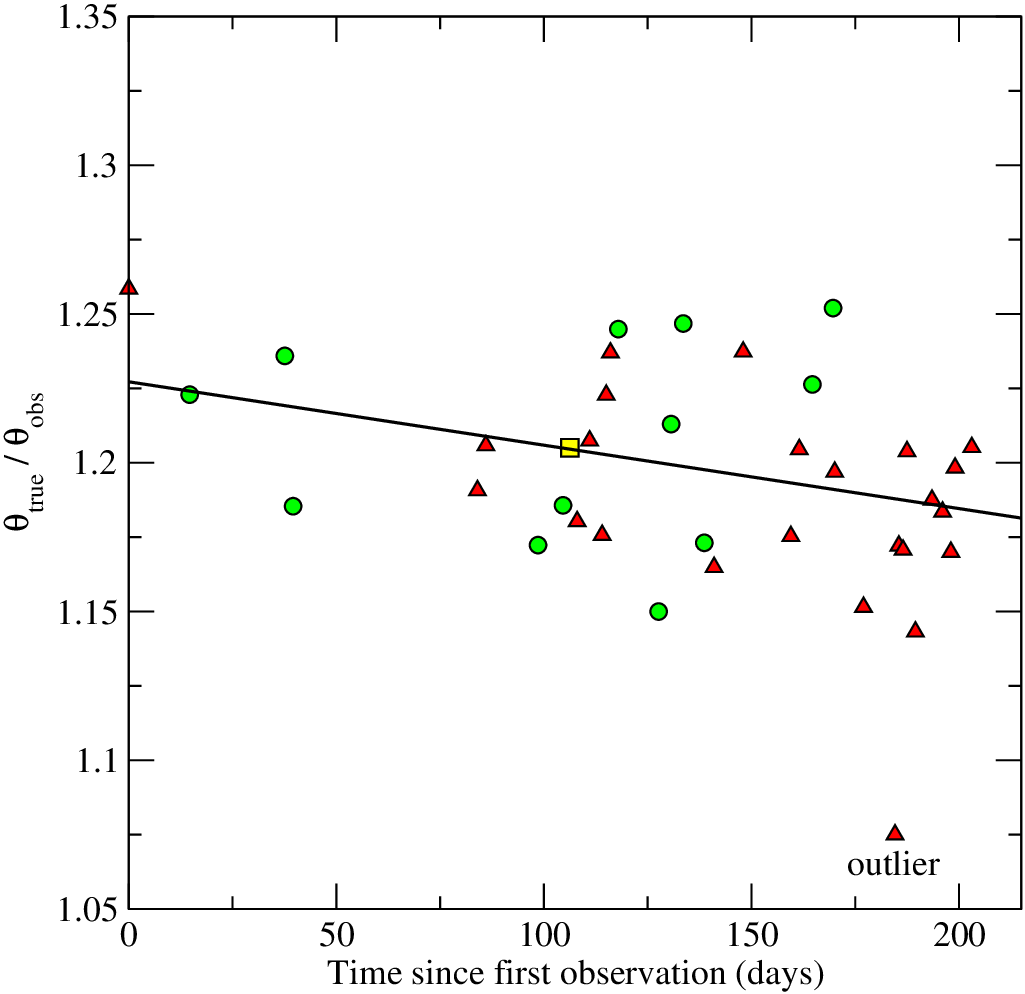]{Ratio of true (geocentric) angular sizes of 
the Moon to the corresponding measured (i.e. uncorrected) values.
\label{ratio}
}

\figcaption[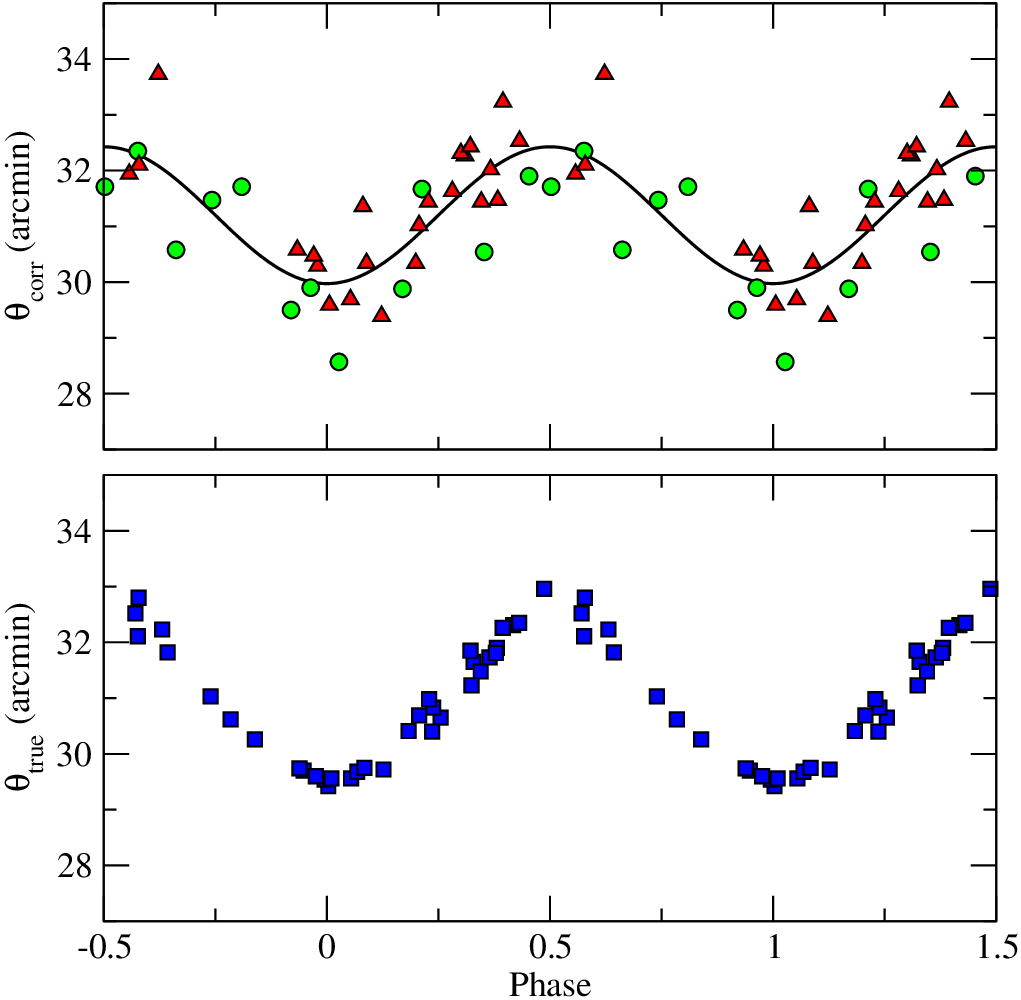]{Corrected values of Moon's angular size and
true geocentric values. \label{ang_diam}
}

\figcaption[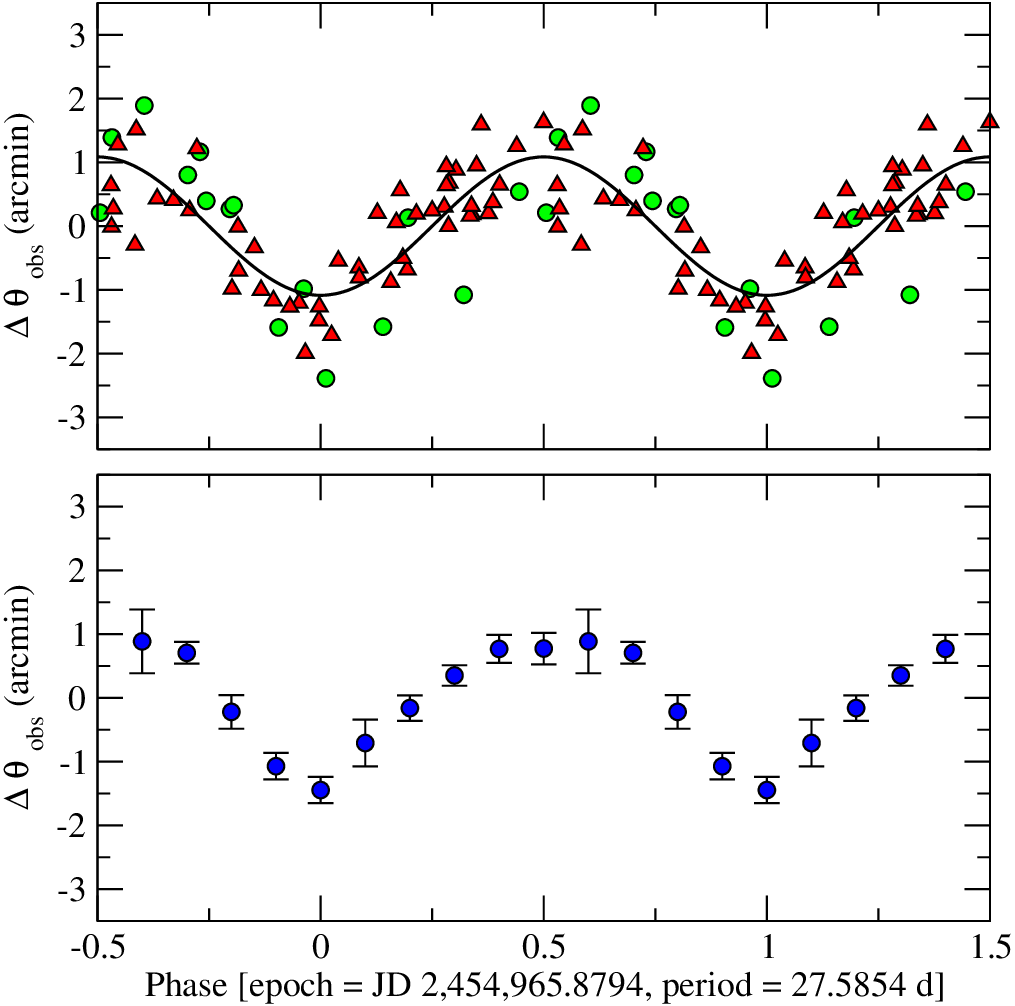]{Phased measures of the variation of the
Moon's angular size, both unbinned and binned.  One
outlier has been eliminated from the analysis.
\label{ang_diam_64}
}

\clearpage

\begin{figure}
\plotone{cross_staff.eps} {Fig. \ref{cross_staff}.  Link to graphic showing a simple Moon
measuring device. 
}
\end{figure}

\begin{figure}
\plotone{ratio.eps}
{Fig. \ref{ratio}. Ratios of the true (geocentric) angular
sizes of the Moon to the corresponding measured (i.e. uncorrected) values.
Data represented by triangles use
observations made during twilight or daytime.   Data
represented by dots use observations made at nighttime when
the observer's pupil was presumably smaller.  The square
represents a calibration done with a 91 mm disk viewed at 10 m,
which matches the true mean angular size of the Moon.  The slope
of the regression line is non-zero at the 2.3-$\sigma$ level,
which is to say it is not statistically significantly different
than zero. 
}
\end{figure}

\begin{figure}
\plotone{ang_diam.eps}
{Fig. \ref{ang_diam}. Angular diameter of the Moon vs. phase
in its cycle from apogee to apogee.  {\em Top}: Corrected
values from 21 April through 10 November 2009.
Data represented by triangles were made during twilight
or daytime.  Data represented by dots were made at nighttime.
The solid line is a single sine wave whose amplitude and mean
value were given by program PERDET.
{\em Bottom}: True geocentric values of the Moon's angular size,
as interpolated from the {\em Astronomical Almanac}.
}
\end{figure}

\begin{figure}
\plotone{ang_diam_64.eps}
{Fig. \ref{ang_diam_64}.  After fitting a fourth order polynomial to
64 uncorrected measures of the Moon's angular diameter, we have determined
the periodicity of the residuals.  {\em Top}: individual
measures.  Triangles represent data taken during twilight and daytime.
Dots represent data taken at night.  {\em Bottom}: data
binned by 0.1 in phase.  The averages are based on 4 to 12 data
points per bin.
}
\end{figure}

\end{document}